\documentclass[conference]{IEEEtran}
\usepackage{calc}

\usepackage{multirow}
\usepackage{cite}
\usepackage{graphicx,subfigure}
\usepackage{psfrag}
\usepackage{amsmath,amssymb}
\usepackage{color}
\usepackage{soul}
\usepackage{algorithm}
\usepackage{algorithmic}

\newcommand{\bit}{\begin{itemize}}
\newcommand{\eit}{\end{itemize}}

\newcommand{\bc}{\begin{center}}
\newcommand{\ec}{\end{center}}

\newcommand{\ba}{\begin{array}}
\newcommand{\ea}{\end{array}}

\newcommand{\beq}{\begin{equation}}
\newcommand{\eeq}{\end{equation}}

\newcommand{\beqn}{\begin{equation*}}
\newcommand{\eeqn}{\end{equation*}}

\newcommand{\bean}{\begin{eqnarray*}}
\newcommand{\eean}{\end{eqnarray*}}
\newcommand{\bea}{\begin{eqnarray}}
\newcommand{\eea}{\end{eqnarray}}

\IEEEoverridecommandlockouts

\begin{document}
\title{Deep Learning for Joint MIMO Detection and Channel Decoding}

\author{Taotao Wang, Lihao Zhang and Soung Chang Liew
	
%

\thanks{T. Wang is with College of Information Engineering, Shenzhen University and the Department of Information Engineering, The Chinese University of Hong Kong (ttwang@szu.edu.cn). Lihao Zhang and Soung Chang Liew are with the Department of Information Engineering, The Chinese University of Hong Kong (zl018@ie.cuhk.edu.hk, soung@ie.cuhk.edu.hk)}
}



\maketitle

\begin{abstract}
We propose a deep-learning approach for the joint MIMO detection and channel decoding problem. Conventional MIMO receivers adopt a model-based approach for MIMO detection and channel decoding in linear or iterative manners. However, due to the complex MIMO signal model, the optimal solution to the joint MIMO detection and channel decoding problem (i.e., the maximum likelihood decoding of the transmitted codewords from the received MIMO signals) is computationally infeasible. As a practical measure, the current model-based MIMO receivers all use suboptimal MIMO decoding methods with affordable computational complexities. This work applies the latest advances in deep learning for the design of MIMO receivers. In particular, we leverage deep neural networks (DNN) with supervised training to solve the joint MIMO detection and channel decoding problem. We show that DNN can be trained to give much better decoding performance than conventional MIMO receivers do. Our simulations show that a DNN implementation consisting of seven hidden layers can outperform conventional model-based linear or iterative receivers. This performance improvement points to a new direction for future MIMO receiver design.
\end{abstract}


\section{Introduction}
Multiple-antenna technology, also known as multiple-input multiple-output (MIMO),  is one of the most important techniques for advanced wireless communications systems. It has already been incorporated into many wireless standards, e.g., 802.11n/ac \cite{802_11ac} and LTE 4G \cite{lte_mimo}. It has also been shown theoretically that MIMO can increase spectrum efficiency linearly with the numbers of transmit and receive antennas \cite{mimo_capacity}. Of much interest are low-complexity MIMO functional units that have good performance.  

A MIMO transmitter transmits multiple data streams, one on each transmit antenna. A MIMO receiver receives a multiplexed copy of the multiple data streams plus noise on each receive antenna. A MIMO detector demultiplexes and decodes the multiplexed data on all the receive antennas into the originally transmitted multiple data streams plus noise and interference.   

To achieve near-capacity performance, advanced channel coding schemes, such as LDPC and polar codes, have been suggested for 5G systems \cite{ldpc,polar}. These channel codes protect the data streams against channel fading, interference, and noise. The output of a MIMO detector consists of a noisy version of the codeword transmitted by the transmitter.  The function of channel decoding is to map the noisy codeword to the original information bits at the transmitter. 

For optimal MIMO decoding, MIMO detection and channel decoding need to be performed in a joint manner. The conventional MIMO decoding schemes all use a model-based approach. However, due to the complex MIMO signal model, the optimal solution to the joint MIMO detection and channel decoding problem (i.e., the maximum likelihood decoding of the transmitted codewords from the received MIMO signals) is computationally infeasible. 

As a practical measure, the current model-based MIMO receivers all use suboptimal MIMO decoding methods with affordable computational complexities. For example, instead of joint MIMO detection and channel decoding,  \cite{linearMIMO1,linearMIMO2,linearMIMO3} proposed to perform MIMO detection and channel decoding sequentially and separately, where MIMO detection is realized by linear equalizations with zero forcing (ZF) or minimum mean square error (MMSE) criteria. By contrast, \cite{iterativeMIMO1,iterativeMIMO2,iterativeMIMO3} proposed to perform MIMO decoding and channel decoding iteratively with soft information exchanges between the two components. Thus, MIMO detection and channel decoding are performed in a joint manner. However, to contain complexity, the original MIMO signal model has been relaxed and replaced by an approximate model (i.e., it separately models the MIMO signal and the channel code). As a result, the solutions are still suboptimal. This leaves a gap for further performance improvement with better MIMO decoder designs. 

To narrow the performance gap, this work applies the latest advances in deep learning for the design of MIMO receivers. In particular, we leverage deep neural networks (DNN) with supervised training to solve the joint MIMO detection and channel decoding problem. We show that DNN can be trained to give much better decoding performance than conventional MIMO receivers do. Our simulations show that a DNN implementation consisting of seven hidden layers can outperform conventional model-based linear or iterative receivers. 

\subsection{Related Work}
Many MIMO detection schemes have been proposed \cite{MIMO_Det}.  Linear MIMO detection can first be used to cancel multiple-antenna interference with low complexities; after that channel decoding is performed  \cite{linearMIMO1,linearMIMO2,linearMIMO3}. In these schemes, linear MIMO detection and channel decoding operate in a sequential manner. Since linear MIMO detection introduces noise amplification and correlation, such sequential linear MIMO detection and channel decoding schemes typically result in large performance loss due to the mismatch between the noise models at the output of the MIMO detector and the input of the channel decoder. 

To enhance the performance of MIMO detection, nonlinear MIMO detectors have also been proposed, e.g., MIMO detectors based on sphere decoding \cite{hardsphere1,hardsphere2,hardsphere3}, semi-definite relaxation \cite{sdr1,sdr2}, and lattice reduction \cite{lra, hardsphere3}. Unfortunately, these nonlinear MIMO detectors can only output hard estimates of channel symbols, making them incompatible with modern channel decoders that require soft input to achieve superior decoding performance. 

Sphere decoding and list decoding algorithms were used for soft MIMO detection \cite{soft_mimo,iterativeMIMO1,iterativeMIMO2,iterativeMIMO3} that produces soft output. This soft information can then be fed to a channel decoding. Moreover, information exchange can be performed iteratively between soft MIMO detection and channel decoding to improve the overall performance of MIMO decoding. Although these iterative MIMO decoding schemes have better performance than the sequential schemes, their solutions are still approximate and suboptimal, due to the mismatch between the noise model of the soft output of the MIMO detector and the assumed noise model at the input of channel decoder. Furthermore, iterative information exchange introduces large decoding latencies.

Unlike the above model-based approaches, \cite{Deep_MIMO} proposed a deep learning approach for MIMO detection. Specifically, the method approximates MIMO detection using deep neural networks (DNN). The method progressively improves the approximation by adjusting the weights of a DNN based on a series of training MIMO signals. Compared with model-based MIMO detection, deep-learning MIMO detection achieves similar detection accuracies with faster detection speed. However, this deep-learning MIMO detection scheme can only perform hard MIMO detection and cannot be combined with a soft channel decoding scheme.

DNN is used to perform channel decoding for the first time in \cite{Deep_decoding}, followed by further work in \cite{Deep_decoding2, Deep_decoding3}. It was shown that DNN channel decoding can approach the MAP performance with lower decoding latency than traditional channel decoding. Work \cite{Deep_linear_codes} employed a neural network constructed by unfolding the factor graph of linear codes to improve the performance of belief propagation decoding when the factor graph of the linear codes contains many samll loops.  Work \cite{Deep_eq} investigated DNN-based joint equalization and channel decoding problem for non-MIMO systems. A survey on the applications of deep learning to wireless systems can be found in \cite{DeepWireless}. 

The remainder of this paper is organized as follows. Section II presents the system model of MIMO systems. Section III reviews the existing model-based MIMO receivers. Section IV presents our deep learning MIMO receiver. Section V provides the simulation results. Finally, Section V concludes the paper.

\section{System Model}

This section presents the system model of MIMO systems and the format of the received MIMO signals. 

Consider a MIMO system where the transmitter is equipped with ${M_T}$ antennas and the receiver is equipped with ${M_R}$ antennas. The channel between each transmit-receive antenna pair is assumed to incur frequency-flat fading and the channel state remains constant within one transmitted packet. We assume ${M_T} \le {M_R}$ and ${M_T}$ parallel data streams are transmitted, one on each transmit antenna.

Figure 1 shows the block diagram of the MIMO transmitter. At the transmitter side, a vector of $K$ information bits, ${\bf{b}} = {\left[ {{b_1},{b_2}, \cdots ,{b_K}} \right]^T}$, is first channel-encoded into a codeword vector ${\bf{c}} = {\left[ {{c_1},{c_2}, \cdots ,{c_N}} \right]^T}$ of length $N = {K \mathord{\left/{\vphantom {K R}} \right.\kern-\nulldelimiterspace} R}$, where $R$ is the code rate. The valid set of codewords is denoted by $\bf C$ and thus ${\bf{c}} \in \bf C$. The coded bits in vector $\bf c$ are modulated to a vector of complex data symbols, ${\bf{\tilde x}} = {\left[ {{x_1},{x_2}, \cdots ,{x_{{N \mathord{\left/	{\vphantom {N B}} \right.
						\kern-\nulldelimiterspace} B}}}} \right]^T}$, where $B$ is the number of code bits per complex data symbol. The modulation constellation is scaled so that  the modulated symbols in ${\bf{\tilde x}}$ have unit average power. Through serial-to-parallel conversion, the vector ${\bf{\tilde x}}$ is partitioned into $L \buildrel \Delta \over = {N \mathord{\left/ {\vphantom {N {\left( {B{M_T}} \right)}}} \right. \kern-\nulldelimiterspace} {\left( {B{M_T}} \right)}}$ consecutive data vectors of length ${M_T}$, $\left\{ {{{\bf{x}}_1},{{\bf{x}}_2}, \cdots ,{{\bf{x}}_L}} \right\}$, i.e., we have ${\bf{\tilde x}} = {\left[ {{\bf{x}}_1^T,{\bf{x}}_2^T, \cdots ,{\bf{x}}_L^T} \right]^T}$.  Then, $L'$ pilot vectors of length ${M_T}$, $\left\{ {{{\bf{p}}_1},{{\bf{p}}_2}, \cdots ,{{\bf{p}}_{L'}}} \right\}$, are prepended to the data vectors $\left\{ {{{\bf{x}}_1},{{\bf{x}}_2}, \cdots ,{{\bf{x}}_L}} \right\}$ to form an ${M_T} \times \left( {L' + L} \right)$ signal matrix ${\bf{X}} = \left[ {{{\bf{X}}_p},{{\bf{X}}_d}} \right]$, where ${{\bf{X}}_d} \buildrel \Delta \over = \left[ {{{\bf{x}}_1},{{\bf{x}}_2}, \cdots ,{{\bf{x}}_L}} \right]$ is the ${M_T} \times L$ data matrix that contains the data vectors, and ${{\bf{X}}_p} \buildrel \Delta \over = \left[ {{{\bf{p}}_1},{{\bf{p}}_2}, \cdots ,{{\bf{p}}_{L'}}} \right]$ is the ${M_T} \times L'$ pilot matrix that contains the pilot vectors. We assume $L' \ge {M_T}$ to facilitate the channel estimation \cite{chaest}. The signal matrix ${\bf{X}}$ represents one transmitted packet. The ${M_T}$ symbols of the  $t$-th column vector in the signal matrix ${\bf{X}}$ are simultaneous transmitted on the ${M_T}$  transmit antennas in the $t$-th time slot.

At the receiver side, the received signals are written into an ${M_R} \times \left( {L' + L} \right)$ matrix, ${\bf{Y}} = \left[ {{{\bf{y}}_1},{{\bf{y}}_2}, \cdots ,{{\bf{y}}_{L' + L}}} \right]$,  where the  $t$-th vector ${{\bf{y}}_t}$ contains the received signals on the ${M_R}$ receive antennas in the  $t$-th time slot.  The received signal matrix can be written as  
\begin{equation}
{\bf{Y}} = \sqrt {\frac{1}{{{M_T}}}} {\bf{HX}} + {\bf{W}}
\end{equation}
where $\bf H$  is an  ${M_R} \times {M_T}$ complex channel matrix with zero-mean and  $\sigma^2$-variance independent complex Gaussian entries, and ${\bf{W}}$ is the ${M_R} \times \left( {L' + L} \right)$
additive white Gaussian noise (AWGN) matrix that has zero-mean and unit-variance independent complex Gaussian entries. We also divide the received signal matrix and the AWGN matrix into two subparts: ${\bf{Y}} = \left[ {{{\bf{Y}}_p},{{\bf{Y}}_d}} \right]$, ${\bf{W}} = \left[ {{{\bf{W}}_p},{{\bf{W}}_d}} \right]$, where ${{\bf{Y}}_p}$ is the  ${M_R} \times L'$ matrix that contains the received signal vectors for the transmitted pilot vectors, ${{\bf{Y}}_d}$  is the  ${M_R} \times L$ matrix that contains the received signal vector for the transmitted data vectors, and ${{\bf{W}}_p}$, ${{\bf{W}}_d}$, are the matrices containing the noise components in ${{\bf{Y}}_p}$, ${{\bf{Y}}_d}$, respectively. The aim of the MIMO receiver is to decode the transmitted information bits in ${\bf{b}}$
from the received signal matrix ${\bf{Y}}$.

For comparison with our proposed MIMO receiver, in Section 3 we review some conventional model-based MIMO receivers.

\begin{figure}[t]
	\centering
	\includegraphics[width=9 cm]{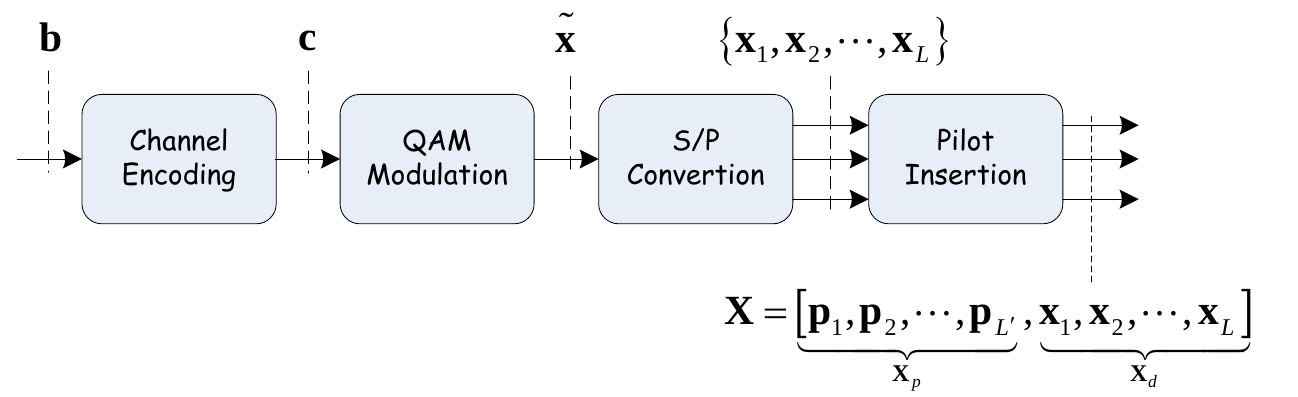}
	\caption{The block diagram for the MIMO transmitter.}
\end{figure}

\section{Model-based MIMO Receivers}

Traditional MIMO receivers have been extensively studied in the literature and implemented in real systems. This section gives a brief overview of these MIMO receivers. 

A symbol-wise optimal MIMO receiver decodes each information bit, ${b_k}$, from the received signal matrix ${{\bf{Y}}_d}$ by minimizing the symbol error probability or equivalently maximizing the \emph{a posteriori probability} (APP):
\begin{equation}
{\hat b_k} = \arg \mathop {\max }\limits_{{b_k}} p\left( {{b_k}\left| {{{\bf{Y}}_d},{\bf{H}},C} \right.} \right)
\end{equation}
where ${\hat b_k}$ denotes the estimate of the information bit ${b_k}$, and $k \in \left\{ {1,2, \cdots ,K} \right\}$. The problem as expressed in (2) is in fact  a \emph{joint MIMO detection and channel decoding} problem, since data symbol detection and the channel decoding are implicitly performed in (2). We point out that joint MIMO detection and channel decoding as in (2) require the knowledge of the channel matrix  ${\bf{H}}$. In practice, the channel matrix is typically estimated from the received pilot signals ${{\bf{Y}}_p}$, e.g., the least square (LS) estimate of the channel matrix is given by: ${\bf{\hat H}} = \sqrt {{M_T}} {{\bf{Y}}_p}{\bf{X}}_p^H{\left( {{{\bf{X}}_p}{\bf{X}}_p^H} \right)^{ - 1}}$\cite{chaest}; then, the channel matrix estimate  ${\bf{\hat H}}$ is substituted back to (2) to replace the real channel matrix ${\bf{H}}$.  

Even with the above approximation which replaces ${\bf{H}}$ by ${\bf{\hat H}}$, the exact computation of  APP, $p\left( {{b_k}\left| {{{\bf{Y}}_d},{\bf{\hat H}},C} \right.} \right)$, is difficult and highly complex. The computation difficulty is due to: i) the correlation among the data symbols introduced by channel encoding; ii) the parallel signal interference caused by the MIMO channel. Therefore, suboptimal MIMO detection and channel decoding schemes with manageable implementation complexities are typically used in practice. We overview two suboptimal schemes in the following. 

\subsection{Linear MIMO Receivers} 

One suboptimal MIMO detection and channel decoding approach is to cancel the parallel signal interference with a linear MIMO detection first and then perform channel decoding next. We refer to this approach  as linear MIMO receivers. For example, the zero-forcing (ZF) detection \cite{linearMIMO1} removes the interference by
\begin{equation}
{{\bf{\tilde Y}}_d} = \sqrt {{M_T}} {\left( {{\bf{ \hat H}}{{\bf{\hat H}}^{{H}}}} \right)^{ - 1}}{{\bf{\hat H}}^H}{{\bf{Y}}_d} = {{\bf{X}}_d} + {{\bf{\tilde W}}_d}
\end{equation}
where ${{\bf{\tilde Y}}_d}$ is the post-cancellation signals and ${{\bf{\tilde W}}_d} = \sqrt {{M_T}} {\left( {{{\bf{H}}_d}{\bf{H}}_d^H} \right)^{ - 1}}{{\bf{W}}_d}$
is the post-cancellation noise. Since parallel signal interference is already removed in (3), the post-cancellation signals, ${{\bf{\tilde Y}}_d}$, can be fed to a traditional channel decoder to recover data symbols. Figure 2 shows the block diagram for this linear MIMO receiver. 

There is no loss of information in (3) since one can get back  ${{\bf{Y}}_d}$ from ${{\bf{\tilde Y}}_d}$. The suboptimality in the linear MIMO decoding arises from the fact that the traditional channel decoder assumes the transformed noise ${{\bf{\tilde W}}_d}$ is white, but it is actually not after the transformation in (3). Although the complexity of this linear MIMO receiver is low, its performance is far from optimal. 

\begin{figure}[t]
	\centering
	\includegraphics[width=9 cm]{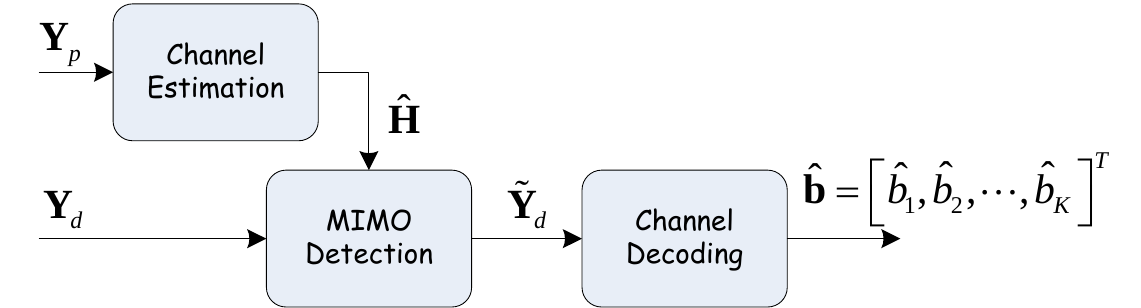}
	\caption{The block diagram for the linear MIMO receiver}
\end{figure}

\subsection{Iterative MIMO Receivers} 

The second MIMO detection and channel decoding approach performs iterative soft-in soft-out MIMO detection and channel decoding. Using ${\bf{\hat H}}$, ${{\bf{Y}}_d}$ and the prior information about the data symbols, a soft MIMO detector computes the extrinsic information about the data symbols \cite{iterativeMIMO1} and delivers the soft information to a soft channel decoder. The soft channel decoder then computes the new extrinsic information about the data symbols and send the computed new extrinsic information back to the soft MIMO detector for further iteration. 

In the next round of iteration, the soft MIMO detector replaces the prior information about the data symbols with the information sent from the soft channel decoder and re-computes its extrinsic information about these data symbols again. Several rounds of such iterations are performed to ensure the convergence of the overall MIMO detection and channel decoding process. We refer to such iterative MIMO detection and channel decoding schemes as iterative MIMO receivers. It yields an approximate solution to the joint MIMO detection and channel decoding problem expressed in (2). 

Figure 3 shows the block diagram for the iterative MIMO receiver. The soft MIMO detection often used is the sphere algorithm \cite{iterativeMIMO3} and the soft channel decoding often used is the belief propagation algorithm. The complexity of the iterative MIMO receiver is much higher than that of the linear MIMO receiver. Although the iterative MIMO receiver has better performance than the linear MIMO receiver does, there is still a large performance gap with respect to the optimal MIMO receiver. Moreover, the iterative information exchange introduces large decoding latency.

\begin{figure}[t]
	\centering
	\includegraphics[width=9 cm]{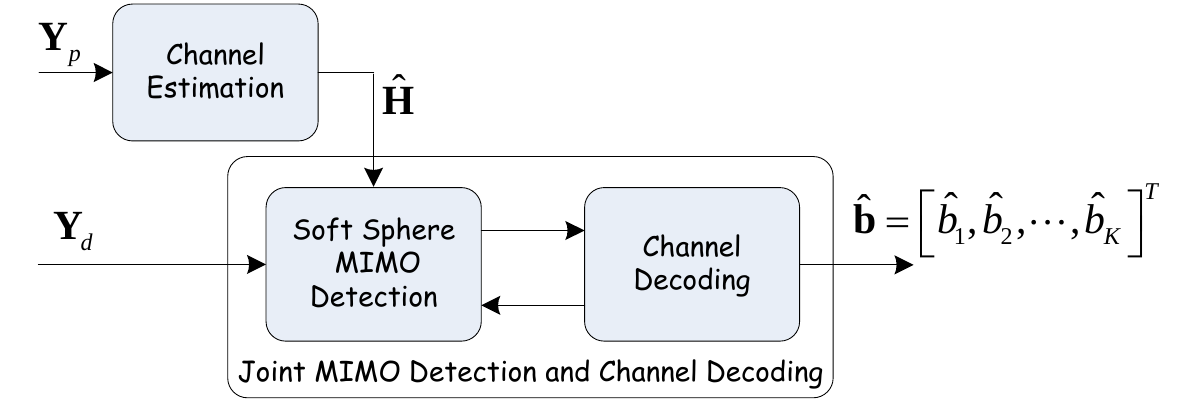}
	\caption{The block diagram for the iterative MIMO receiver.}
\end{figure}

\section{Deep-Learning MIMO Receivers} 

We propose to employ deep neural networks (DNN) to solve the joint MIMO detection and channel decoding problem stated in (2) with the goal of improving performance. The DNNs are trained under the framework of supervised learning.

We consider the training of DNN at the MIMO receiver after the channel matrix estimate ${\bf{\hat H}}$  has already computed from the received pilot signals. Using this channel matrix estimate ${\bf{\hat H}}$ at the MIMO receiver, we generate a set of training signals to train a DNN to solve the joint MIMO detection and channel decoding problem (2) under the framework of supervised learning. The training and deployment framework of DNN for MIMO is illustrated in Figure 4. We describe the associated procedures in the following.

The receiver generates the training data by calling a functional block that mimics the operation at the MIMO transmitter. Specifically, for training purposes, the receiver randomly generates many length-$K$ binary vectors, ${{\bf{b}}^{\left( i \right)}}$, $i = 1,2, \cdots ,Z$. Each binary vector  ${{\bf{b}}^{\left( i \right)}}$ is transformed into a data matrix ${\bf{X}}_d^{\left( i \right)}$ using the functional block of the MIMO transmitter as described in Section II. Then, with the channel matrix estimate ${\bf{\hat H}}$ given by the channel estimator, the receiver generates a training signal by multiplying ${\bf{\hat H}}$ with ${\bf{X}}_d^{\left( i \right)}$ followed by adding AWGN: 
$${\bf{Y}}_d^{\left( i \right)} = \sqrt {{1 \mathord{\left/
			{\vphantom {1 {{M_T}}}} \right.
			\kern-\nulldelimiterspace} {{M_T}}}} {\bf{\hat HX}}_d^{\left( i \right)} + {\bf{W}}_d^{\left( i \right)}$$
where  ${\bf{Y}}_d^{\left( i \right)}$ is the $i$-th training signal and ${\bf{W}}_d^{\left( i \right)}$ is the corresponding generated AWGN. The training set is given by $D\left( {{\bf{\hat H}}} \right) = \left\{ {{\bf{Y}}_d^{\left( i \right)},{{\bf{b}}^{\left( i \right)}}\left| {{\bf{\hat H}}} \right.} \right\}_{i = 1}^Z$, where ${\bf{Y}}_d^{\left( i \right)}$ is the  $i$-th training signal and ${{\bf{b}}^{\left( i \right)}}$ is the corresponding label for  ${\bf{Y}}_d^{\left( i \right)}$. We emphasize that the training set is dependent on the channel matrix estimate ${\bf{\hat H}}$. 

\begin{figure*}[t]
	\centering
	\includegraphics[width=15cm]{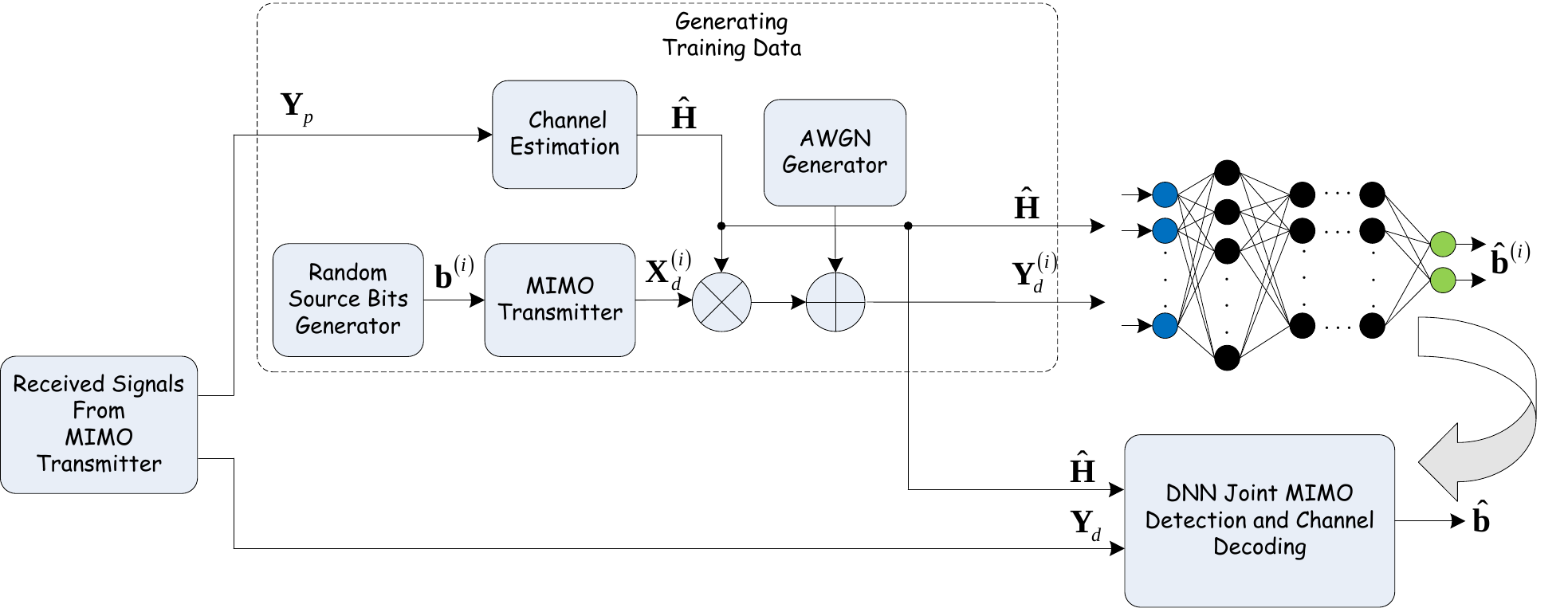}
	\caption{The training and deployment framework of DNN for MIMO.}
\end{figure*}

We use the generated training set to train a DNN, ${f_{{\theta }}}\left(  \cdot  \right)$, that approximates the solution to problem (2), where  ${{\theta }}$ is the set containing all the weights of the edges in the DNN.  When we feed the training signals  $\left\{ {{\bf{Y}}_d^{\left( i \right)}} \right\}_{i = 1}^Z$ to the inputs of the DNN, we also feed the channel matrix estimate ${\bf{\hat H}}$ to the DNN (as illustrated in Figure 4). We optimize the DNN weights by miming the cross entropy loss function \cite{Goodfellow}: 
\begin{equation}
\begin{array}{l}
L\left( {\theta ,D\left( {{\bf{\hat H}}} \right)} \right) \\ 
= \frac{1}{{ZK}}\sum\limits_{i = 1}^Z {\sum\limits_{k = 1}^K {\left[ {b_k^{\left( i \right)}\ln \left( {\hat b_k^{\left( i \right)}} \right) + \left( {1 - b_k^{\left( i \right)}} \right)\ln \left( {1 - \hat b_k^{\left( i \right)}} \right)} \right]} }  \\ 
\end{array}
\end{equation}
where $b_k^{\left( i \right)} \in \left\{ {0,1} \right\}$ is the $k$-th target information bit of the $i$-th label vector ${{\bf{b}}^{\left( i \right)}}$, $\hat b_k^{\left( i \right)}$ is the soft estimate of $b_k^{\left( i \right)} \in \left\{ {0,1} \right\}$ given by the DNN. The training algorithm used to minimize (4) for DNN is the so called stochastic gradient descent (SGD) algorithm \cite{Goodfellow}. After the training is finished, the weights of the DNN are fixed to ${{\hat \theta }}$ and we can use the trained DNN  ${f_{{{\hat \theta }}}}\left( \cdot \right)$ to decode the received signals as ${\bf{\hat b}} = {f_{{{\hat \theta }}}}\left( {{{\bf{Y}}_d}} \right)$.  We have the following remarks on this DNN for MIMO:

\begin{itemize}[leftmargin=*,labelsep=5.8mm]
	
	\item The variables of interests to the DNN are the data symbols in ${\bf{X}}_d^{\left( i \right)}$. The size of the variable space is thus ${2^K}$, where $K$ is the length of ${{\bf{b}}^{\left( i \right)}}$  (Note that we have the one-to-one mapping: ${{\bf{b}}^{\left( i \right)}} \to {\bf{X}}_d^{\left( i \right)}$). According to the results shown in \cite{Deep_decoding}, if the DNN can see all possible codewords, the decoding performance of the DNN is the best. Like the investigation in \cite{Deep_decoding}, we also adopt short codes and train the DNN with all different codewords. 
	
	
	\item The training of the  DNN is quite time-consuming. Therefore, the training procedure will introduce a large decoding latency and it cannot be deployed for applications with stringent latency requirements, such as voice transmissions; it is, however, suitable for data transmissions with relaxed latency requirements.

	
\end{itemize}

\section{Simulation Results}

In this section, we present simulation results for the evaluation of the proposed DNN MIMO receiver. The modulations used are BPSK and QPSK. The channel code used is the polar code \cite{polar} with code rate $1/2$. We assume that  that each packet consists of $K=16$ bits in the simulations. The adoption of the  short packet length is due to the exponential training complexity when DNN is used to perform channel decoding \cite{Deep_decoding}.\footnote{The extension of extend DNN channel decoding to long packet length can follow the solution of  \cite{Deep_decoding3}. We will consider how to incorporate the solution of  \cite{Deep_decoding3} into our DNN joint MIMO detection and channel decoding scheme in future work.}  Packets of short length are of interest in some practical systems such as the internet of things (IoT). After channel encoding and modulation, $K=16$ information bits are transformed to 32 BPSK symbols or 16 QPSK symbols. Our simulations assume MIMO matrices of dimensions $M_R \times M_T = 2\times 2$, $4 \times 4$ and $8 \times 8$.

We implement a DNN consisting of one input layer, six hidden layers and one output layer using the deep-learning software toolkit of Keras. The nonlinear activation function at the neurons of the input layer and the hidden layers is the Rectified linear unit (ReLu) function \cite{Goodfellow}. The input layer is a densely-connected layer. Each hidden layer is a densely-connected layer with batch normalization (BN) operations before the operations by ReLu. The output layer is a densely-connected layer with the sigmoid activation functions. The architecture of the DNN is illustrated in Figure 5. We train our NN over several “epochs”. In each epoch, the gradient of the loss function is computed over the entire training set using Adam, a method for stochastic gradient descent optimization \cite{Adam}. Our training set contains all different $2^{K}$ codewords, $K$ is the length of information bits. Setting the number of learning epochs to $10^{5}$, we train the DNN with datasets of different training SNRs (from 0 dB to 6 dB). After the training is finished, the trained DNN is used to decode the received MIMO signals.

\begin{figure}[h]
	\centering
	\includegraphics[width=4 cm, angle=90]{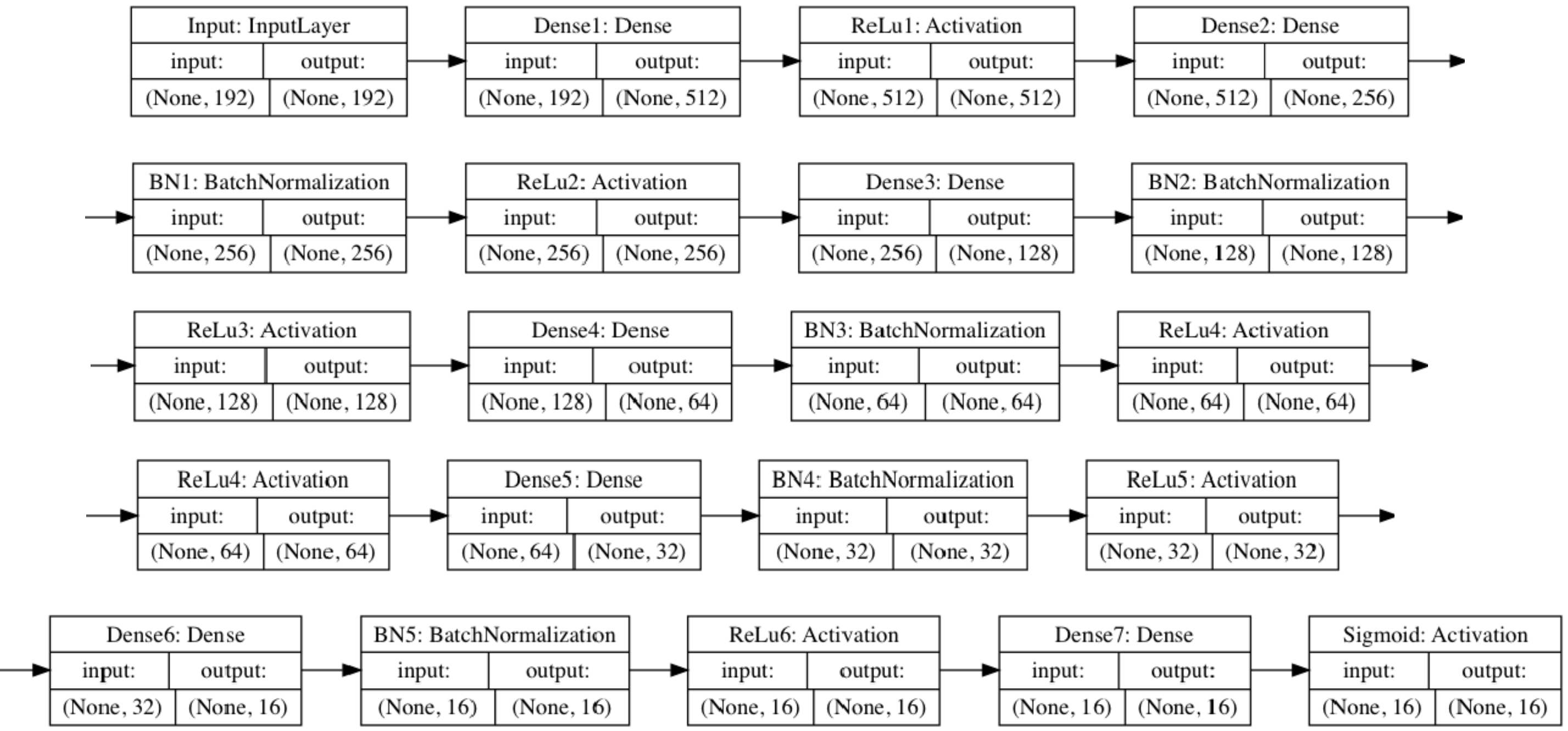}
	\caption{The architecture of the adopted DNN consisting of six hidden layers with 512, 356, 128, 64, 32, and 16 neurons respectively. }
\end{figure}  

For comparison, we treat the following two traditional MIMO receivers as our benchmarks: i) the linear MIMO receiver that employs ZF MIMO detection followed by the MAP polar decoding of \cite{polar}, ii) the iterative MIMO receiver that iterates between the sphere MIMO detection of \cite{iterativeMIMO3} and the MAP polar decoding of \cite{polar}. We investigate the performance of MIMO receivers with perfect knowledge as well as with imperfect knowledge of the channel matrix. For the latter, we assume LS estimation \cite{chaest} is used to estimate the channel matrix. For a fixed SNR, we evaluate the average BER results of the MIMO receivers over 100 different MIMO channel realizations.

Figure 6 and Figure 7 show the BER of the MIMO receivers with perfect knowledge of the MIMO channel matrix for BPSK and QPSK, respectively. We can observe that our DNN MIMO receiver can indeed outperform the linear and iterative MIMO receivers in terms of BER. For example, the DNN MIMO receiver has around 1 dB and 3.5 dB SNR gain over the linear and iterative MIMO receivers, respectively, at the BER of $10^{-4}$ for BPSK and $8 \times 8$ MIMO channels. 

Figure 8 and Figure 9 show the BER of the MIMO receivers with imperfect knowledge of the MIMO channel matrix for BPSK and QPSK, respectively. For the channel matrix estimation, we place a Hadamard matrix at the beginning of the packets as pilots and use the LS estimation based on the received pilots to estimate the channel matrix at the receivers. In general, the performance trend for the cases of perfect and imperfect channel estimates are the same. The only difference between them is that for the cases of imperfect channel estimates, the gain obtained by our DNN MIMO receiver is even larger. For example, the DNN MIMO receiver now has around 2 dB and 10 dB SNR gain over the linear and iterative MIMO receivers at the BER of $10^{-4}$ for BPSK and $8 \times 8$ MIMO channels.


\section{Conclusions}

This work used a deep-learning tool, deep neural network, to develop a new solution to the problem of joint MIMO detection and channel decoding. Conventional MIMO receivers perform MIMO detection and channel decoding in a sequential or an iterative manner. The algorithms of these conventional MIMO receivers relax the signal model of coded MIMO. As a result, they are suboptimal solutions to the joint MIMO detection and channel decoding problem, leaving the possibility for further improvement. Our deep learning solution uses a DNN for joint MIMO detection and channel decoding under the framework of supervised learning. The deep-learning MIMO receiver  does not separate the MIMO detection and channel decoding into two parts and does not perform sequential or iterative operations on them. It treats the MIMO detection and channel decoding as a joint decoding process and employs a single DNN to approximate the joint decoding process. This joint process improves the overall decoding performance. In our simulations, we trained a DNN consisting of six hidden layers to decode MIMO signals. The simulation results demonstrate notable gains obtained by our deep-learning MIMO receiver over the conventional linear and iterative MIMO receivers. 

A drawback of the current proposed deep-learning MIMO receiver is that the DNN needs to be trained for each different channel matrix, introducing a large decoding latency. In general, to train the same DNN for MIMO decoding with different channel matrices is challenging, since the space of all possible channel matrices is huge. It is impossible to let the DNN see all the channel realizations. In \cite{Deep_MIMO}, a scheme to construct one DNN for MIMO detection with different channel matrices is given. However, it is not clear how to extend the associated DNN to solve the problem of joint MIMO detection and channel decoding. A DNN for joint MIMO detection and channel decoding that can handle different channel matrices with one training (i.e., no need to train and readjust the weights in the DNN for each different channel matrix) is an interesting direction for further investigations.  

\begin{figure}[t]
	\centering
	\includegraphics[width=9 cm]{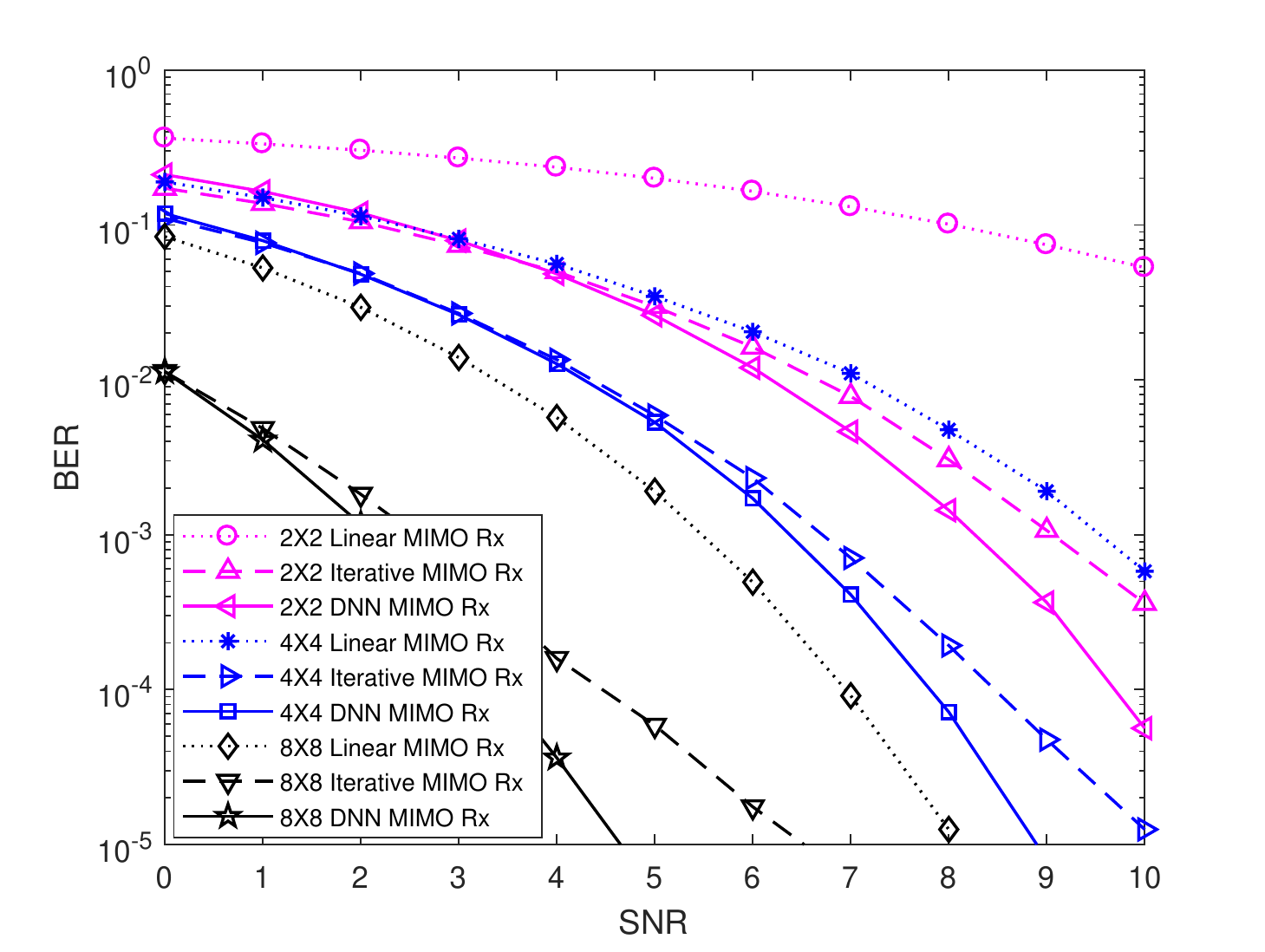}
	\caption{BER of the MIMO receivers with  perfect knowledge of the MIMO channel matrix for BPSK.}
\end{figure}

\begin{figure}[t]
	\centering
	\includegraphics[width=9 cm]{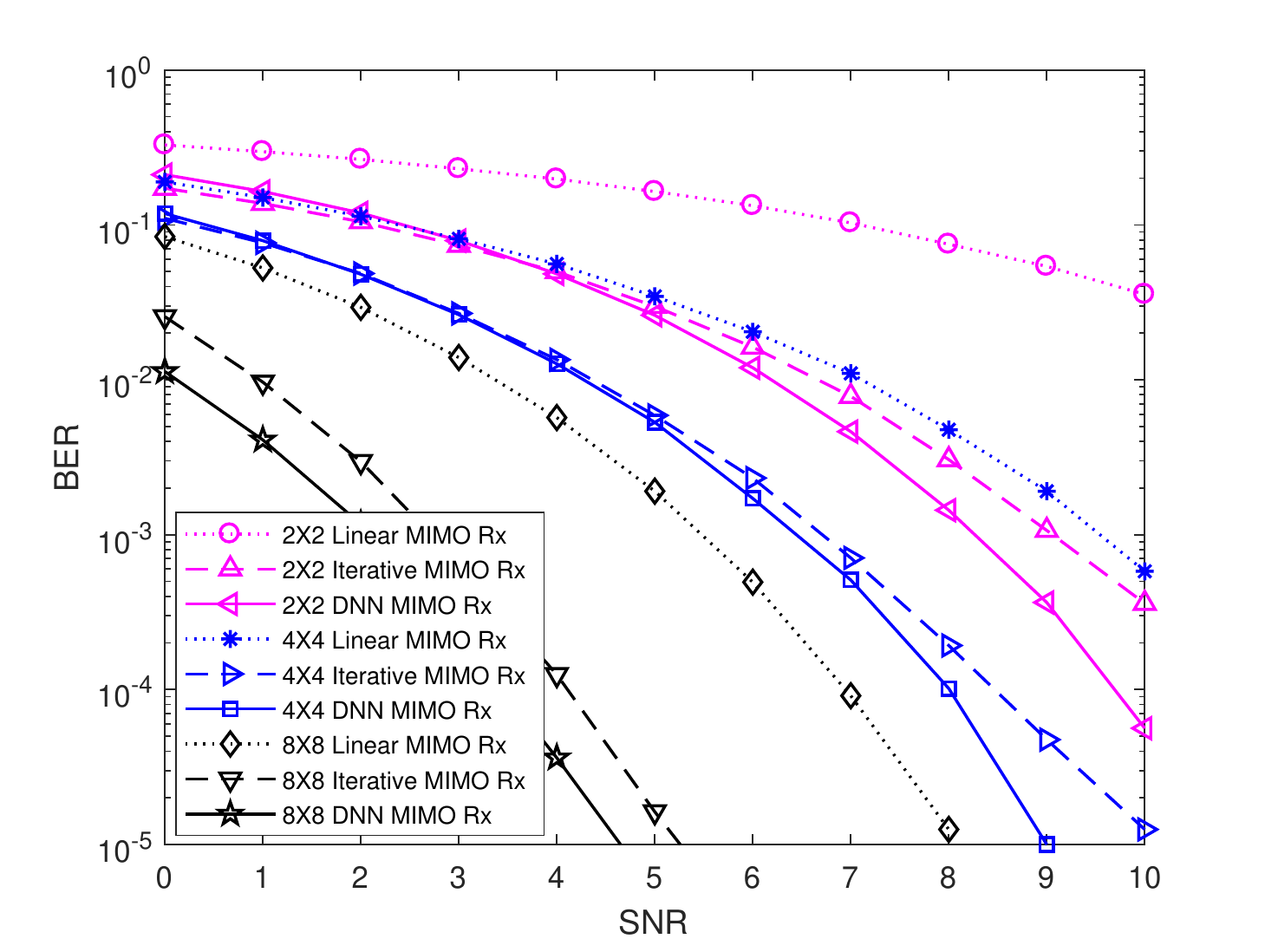}
	\caption{BER of MIMO receivers with  the perfect knowledge of the MIMO channel matrix for QPSK.}
\end{figure}

\begin{figure}[t]
	\centering
	\includegraphics[width=9 cm]{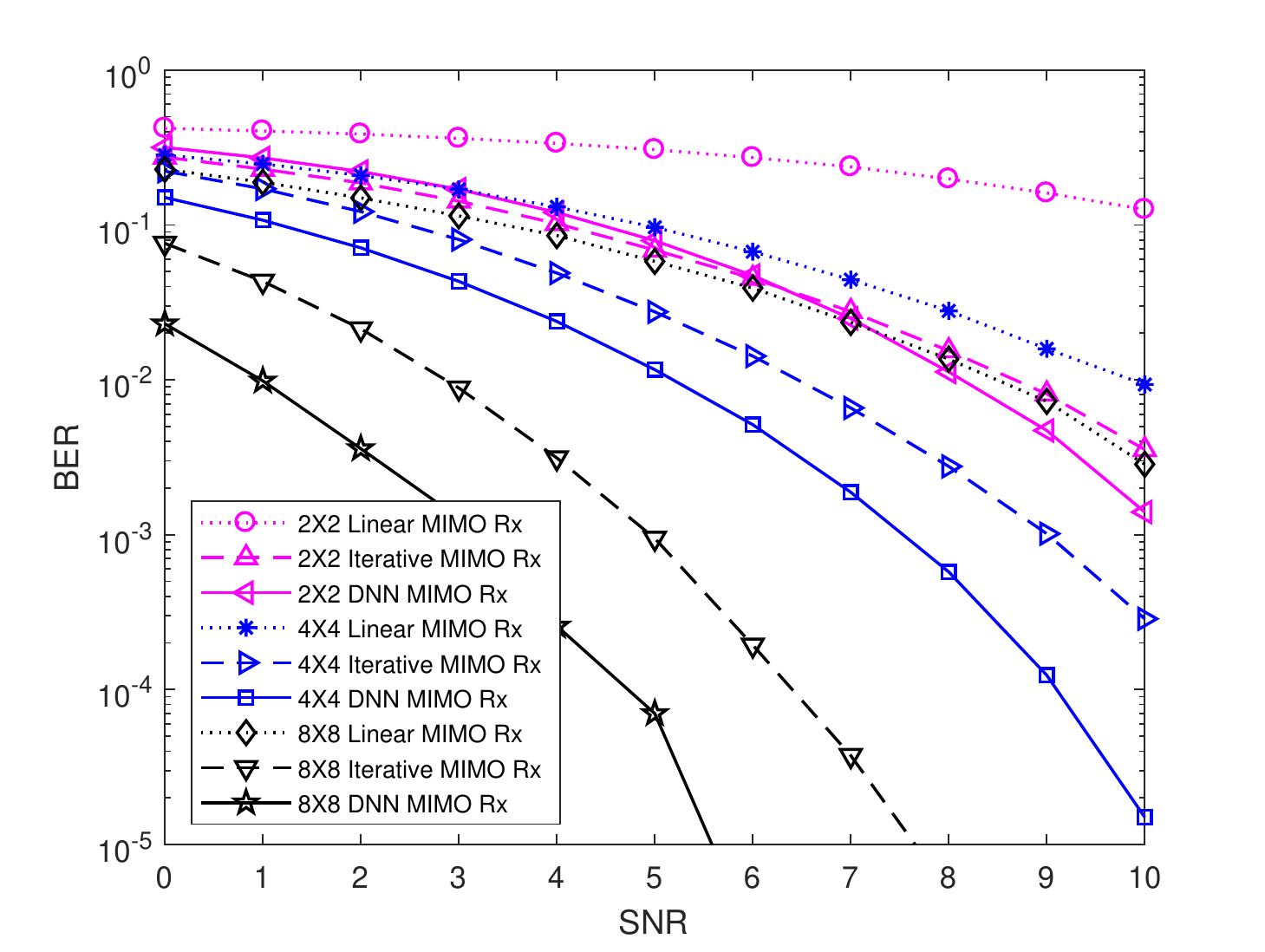}
	\caption{BER of MIMO receivers with imperfect knowledge of the MIMO channel matrix for BPSK.}
\end{figure}

\begin{figure}[t]
	\centering
	\includegraphics[width=9 cm]{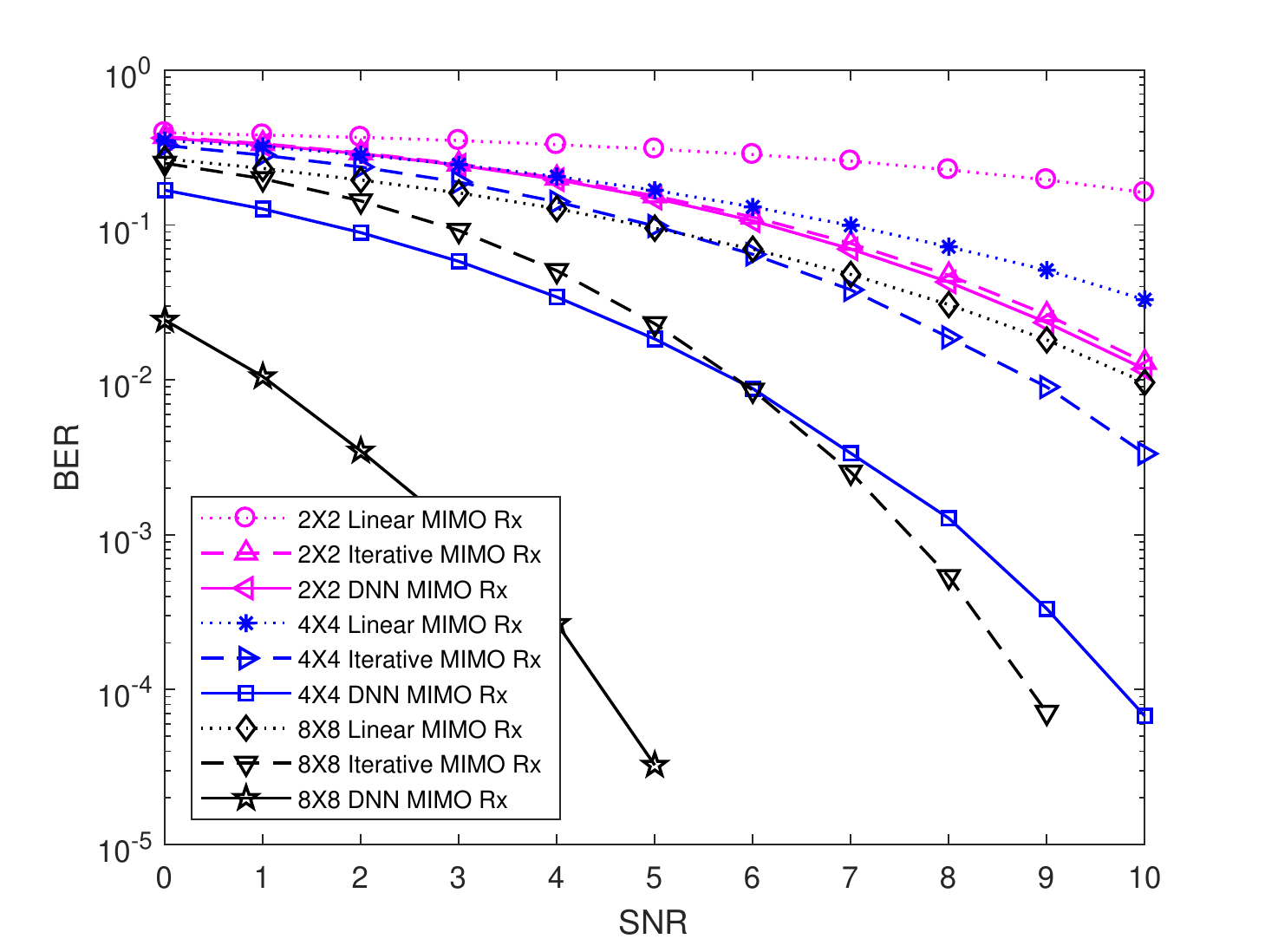}
	\caption{BER of MIMO receivers with imperfect  knowledge of the MIMO channel matrix for QPSK.}
\end{figure}

\ifCLASSOPTIONcaptionsoff
\newpage
\fi

\bibliographystyle{IEEEtran}

\bibliography{database}

\begin{thebibliography}{10}
\providecommand{\url}[1]{#1}
\csname url@samestyle\endcsname
\providecommand{\newblock}{\relax}
\providecommand{\bibinfo}[2]{#2}
\providecommand{\BIBentrySTDinterwordspacing}{\spaceskip=0pt\relax}
\providecommand{\BIBentryALTinterwordstretchfactor}{4}
\providecommand{\BIBentryALTinterwordspacing}{\spaceskip=\fontdimen2\font plus
\BIBentryALTinterwordstretchfactor\fontdimen3\font minus
  \fontdimen4\font\relax}
\providecommand{\BIBforeignlanguage}[2]{{%
\expandafter\ifx\csname l@#1\endcsname\relax
\typeout{** WARNING: IEEEtran.bst: No hyphenation pattern has been}%
\typeout{** loaded for the language `#1'. Using the pattern for}%
\typeout{** the default language instead.}%
\else
\language=\csname l@#1\endcsname
\fi
#2}}
\providecommand{\BIBdecl}{\relax}
\BIBdecl

\bibitem{802_11ac}
O.~Bejarano, E.~W. Knightly, and M.~Park, ``Ieee 802.11 ac: from channelization
  to multi-user mimo,'' \emph{IEEE Communications Magazine}, vol.~51, no.~10,
  pp. 84--90, 2013.

\bibitem{lte_mimo}
A.~Ghosh and R.~Ratasuk, \emph{Essentials of lte and lte-a}.\hskip 1em plus
  0.5em minus 0.4em\relax Cambridge University Press, 2011.

\bibitem{mimo_capacity}
A.~Goldsmith, S.~A. Jafar, N.~Jindal, and S.~Vishwanath, ``Capacity limits of
  mimo channels,'' \emph{IEEE Journal on selected areas in Communications},
  vol.~21, no.~5, pp. 684--702, 2003.

\bibitem{ldpc}
T.~Richardson and S.~Kudekar, ``Design of low-density parity check codes for 5g
  new radio,'' \emph{IEEE Communications Magazine}, vol.~56, no.~3, pp. 28--34,
  2018.

\bibitem{polar}
E.~Arikan, ``Channel polarization: A method for constructing capacity-achieving
  codes for symmetric binary-input memoryless channels,'' \emph{IEEE
  Transactions on Information Theory}, vol.~55, no.~7, pp. 3051--3073, 2009.

\bibitem{linearMIMO1}
T.~Haustein, C.~Von~Helmolt, E.~Jorswieck, V.~Jungnickel, and V.~Pohl,
  ``Performance of mimo systems with channel inversion,'' in \emph{Vehicular
  Technology Conference, 2002. VTC Spring 2002. IEEE 55th}, vol.~1.\hskip 1em
  plus 0.5em minus 0.4em\relax IEEE, 2002, pp. 35--39.

\bibitem{linearMIMO2}
K.~R. Kumar, G.~Caire, and A.~L. Moustakas, ``Asymptotic performance of linear
  receivers in mimo fading channels,'' \emph{arXiv preprint arXiv:0810.0883},
  2008.

\bibitem{linearMIMO3}
A.~Hedayat and A.~Nosratinia, ``Outage and diversity of linear receivers in
  flat-fading mimo channels,'' \emph{IEEE Transactions on Signal Processing},
  vol.~55, no.~12, pp. 5868--5873, 2007.

\bibitem{iterativeMIMO1}
B.~M. Hochwald and S.~Ten~Brink, ``Achieving near-capacity on a
  multiple-antenna channel,'' \emph{IEEE transactions on communications},
  vol.~51, no.~3, pp. 389--399, 2003.

\bibitem{iterativeMIMO2}
E.~Witte, F.~Borlenghi, G.~Ascheid, R.~Leupers, and H.~Meyr, ``A scalable vlsi
  architecture for soft-input soft-output depth-first sphere decoding,''
  \emph{IEEE Transactions on Circuits and Systems II: Express Briefs, 2010}.

\bibitem{iterativeMIMO3}
C.~Studer and H.~Bolcskei, ``Soft--input soft--output single tree-search sphere
  decoding,'' \emph{IEEE Transactions on Information Theory}, vol.~56, no.~10,
  pp. 4827--4842, 2010.

\bibitem{MIMO_Det}
E.~G. Larsson, ``Mimo detection methods: How they work [lecture notes],''
  \emph{IEEE signal processing magazine}, vol.~26, no.~3, 2009.

\bibitem{hardsphere1}
E.~Viterbo and J.~Boutros, ``A universal lattice code decoder for fading
  channels,'' \emph{IEEE Transactions on Information theory}, vol.~45, no.~5,
  pp. 1639--1642, 1999.

\bibitem{hardsphere2}
L.~G. Barbero and J.~S. Thompson, ``Fixing the complexity of the sphere decoder
  for mimo detection,'' \emph{IEEE Transactions on Wireless Communications},
  vol.~7, no.~6, 2008.

\bibitem{hardsphere3}
E.~Agrell, T.~Eriksson, A.~Vardy, and K.~Zeger, ``Closest point search in
  lattices,'' \emph{IEEE transactions on information theory}, vol.~48, no.~8,
  pp. 2201--2214, 2002.

\bibitem{sdr1}
P.~H. Tan and L.~K. Rasmussen, ``The application of semidefinite programming
  for detection in cdma,'' \emph{IEEE journal on selected areas in
  communications}, vol.~19, no.~8, pp. 1442--1449, 2001.

\bibitem{sdr2}
B.~Steingrimsson, Z.-Q. Luo, and K.~M. Wong, ``Soft quasi-maximum-likelihood
  detection for multiple-antenna wireless channels,'' \emph{IEEE Transactions
  on Signal Processing}, vol.~51, no.~11, pp. 2710--2719, 2003.

\bibitem{lra}
C.~Windpassinger and R.~F. Fischer, ``Low-complexity near-maximum-likelihood
  detection and precoding for mimo systems using lattice reduction,'' in
  \emph{Information Theory Workshop, 2003. Proceedings. 2003 IEEE}.\hskip 1em
  plus 0.5em minus 0.4em\relax IEEE, 2003, pp. 345--348.

\bibitem{soft_mimo}
E.~G. Larsson and J.~Jalden, ``Fixed-complexity soft mimo detection via partial
  marginalization,'' \emph{IEEE transactions on Signal Processing}, vol.~56,
  no.~8, pp. 3397--3407, 2008.

\bibitem{Deep_MIMO}
N.~Samuel, T.~Diskin, and A.~Wiesel, ``Deep mimo detection,'' in \emph{IEEE
  18th International Workshop on Signal Processing Advances in Wireless
  Communications (SPAWC)}.\hskip 1em plus 0.5em minus 0.4em\relax IEEE, 2017,
  pp. 1--5.

\bibitem{Deep_decoding}
T.~Gruber, S.~Cammerer, J.~Hoydis, and S.~ten Brink, ``On deep learning-based
  channel decoding,'' in \emph{Information Sciences and Systems (CISS), 2017
  51st Annual Conference on}.\hskip 1em plus 0.5em minus 0.4em\relax IEEE,
  2017, pp. 1--6.

\bibitem{Deep_decoding2}
J.~Seo, J.~Lee, and K.~Kim, ``Decoding of polar code by using deep feed-forward
  neural networks,'' in \emph{2018 International Conference on Computing,
  Networking and Communications (ICNC)}.\hskip 1em plus 0.5em minus 0.4em\relax
  IEEE, 2018, pp. 238--242.

\bibitem{Deep_decoding3}
S.~Cammerer, T.~Gruber, J.~Hoydis, and S.~ten Brink, ``Scaling deep
  learning-based decoding of polar codes via partitioning,'' in \emph{GLOBECOM
  2017-2017 IEEE Global Communications Conference}.\hskip 1em plus 0.5em minus
  0.4em\relax IEEE, 2017, pp. 1--6.

\bibitem{Deep_linear_codes}
E.~Nachmani, E.~Marciano, L.~Lugosch, W.~J. Gross, D.~Burshtein, and
  Y.~Be’ery, ``Deep learning methods for improved decoding of linear codes,''
  \emph{IEEE Journal of Selected Topics in Signal Processing}, vol.~12, no.~1,
  pp. 119--131, 2018.

\bibitem{Deep_eq}
H.~Ye and G.~Y. Li, ``Initial results on deep learning for joint channel
  equalization and decoding,'' in \emph{Vehicular Technology Conference
  (VTC-Fall), 2017 IEEE 86th}.\hskip 1em plus 0.5em minus 0.4em\relax IEEE,
  2017, pp. 1--5.

\bibitem{DeepWireless}
T.~Wang, C.-K. Wen, H.~Wang, F.~Gao, T.~Jiang, and S.~Jin, ``Deep learning for
  wireless physical layer: Opportunities and challenges,'' \emph{China
  Communications}, vol.~14, no.~11, pp. 92--111, 2017.

\bibitem{chaest}
M.~Biguesh and A.~B. Gershman, ``Training-based mimo channel estimation: a
  study of estimator tradeoffs and optimal training signals,'' \emph{IEEE
  transactions on signal processing}, vol.~54, no.~3, pp. 884--893, 2006.

\bibitem{Goodfellow}
Y.~LeCun, Y.~Bengio, and G.~Hinton, ``Deep learning,'' \emph{nature}, vol. 521,
  no. 7553, p. 436, 2015.

\bibitem{Adam}
D.~P. Kingma and J.~Ba, ``Adam: A method for stochastic optimization,''
  \emph{arXiv preprint arXiv:1412.6980}, 2014.

\end{thebibliography}

\end{document}